\documentclass{ICHD2024}

\title{Experimental study of fish-like bodies with passive tail and tunable stiffness}
\author{\underline{L. Padovani}, G. Manduca,  D.Paniccia, G. Graziani, R. Piva and C. Lugni}
\address{%

}

\usepackage{graphicx}
\usepackage{subfigure} 
\usepackage{amsmath} 
\usepackage{subcaption}
\usepackage{appendix}
\begin{document}

\maketitle

\begin{center}
\textbf{ABSTRACT}
\end{center}
Scombrid fishes and tuna are efficient swimmers capable of maximizing performance to escape predators and save energy during long journeys. A key aspect in achieving these goals is the flexibility of the tail, which the fish optimizes during swimming. Though, the robotic counterparts, although highly efficient, have partially investigated the importance of flexibility. We have designed and tested a fish-like robotic platform (of 30 $cm$ in length) to quantify performance with a tail made flexible through a torsional spring placed at the peduncle. Body kinematics, forces, and power have been measured and compared with real fish. The platform can vary its frequency between 1 and 3 $Hz$, reaching self-propulsion conditions with speed over 1 $BL/s$ and Strouhal number in the optimal range. We show that changing the frequency of the robot can influence the thrust and power achieved by the fish-like robot. Furthermore, by using appropriately tuned stiffness, the robot deforms in accordance with the travelling wave mechanism, which has been revealed to be the actual motion of real fish. These findings demonstrate the potential of tuning the stiffness in fish swimming and offer a basis for investigating fish-like flexibility in bio-inspired underwater vehicles.
\section{Introduction}
The necessity to explore locations beyond human reach has driven the scientific community to develop underwater vehicles, which have progressively evolved over the past decades towards bio-inspired configurations. A multidisciplinary approach has led to a sequence of robotic platforms mimicking fish-like bodies with the aim to enhance their swimming performance (\cite{article0}, \cite{article1}). Much remains to be investigated since current systems, though capable of achieving very high speed (\cite{article2}), are not able to reach an efficiency comparable to their natural counterparts.  Recently some of these models have started to investigate the adoption of a passive tail connected to the fish-like body by a torsional spring of given stiffness, to analyze its positive effect on the overall performance. Some experimental studies (see \cite{article3}) have illustrated how to exploit the presence of resonant phenomena, which are able to increase, for certain values of the ruling parameters, the stride length in self-propelled swimming through the use of a tunable stiffness. Theory (\cite{article4}), simulations (\cite{article5}), and experiments (\cite{article6}-\cite{article8}) have demonstrated how flexibility can enhance the performance of fish-like bodies. Mechanisms for adjusting stiffness already exist. Robots can adjust stiffness by changing passive stiffness elements or by utilizing adjustable springs (\cite{article9}-\cite{article12}). New solutions for tunable stiffness have also been proposed using soft actuators and artificial tendons (\cite{article13}-\cite{article14}). \\
Undulatory swimmers are indeed animals that generate a traveling wave along their body or propulsive fins to push fluid backward. They can adjust their stiffness while swimming through active muscle tension. As a consequence, it is straightforward to take inspiration from real models to create bio-robotic platforms capable of achieving performance similar to that of real fish. Tunas and mackerels are excellent biological equivalents for studying high-performance swimming behaviours. They are characterized by high-aspect ratio tails and narrow peduncles, which are used passively to swim quickly and/or efficiently over long distances.
For these reasons, we have decided to consider in our experiments a fish-like body with passive tail under an incoming flow to explore the role of the leading parameters including the velocity of incoming flow, the frequency of the given heave motion and the stiffness of the tail spring. An experimental model of a robotic fish has been used in a small water tunnel to evaluate its behaviour for different values of the above parameters and their potential impact on the efficiency by measuring the interacting forces.

\section{Materials and methods}
\begin{figure}[h!]
\centering
\includegraphics[width=\linewidth,clip]
{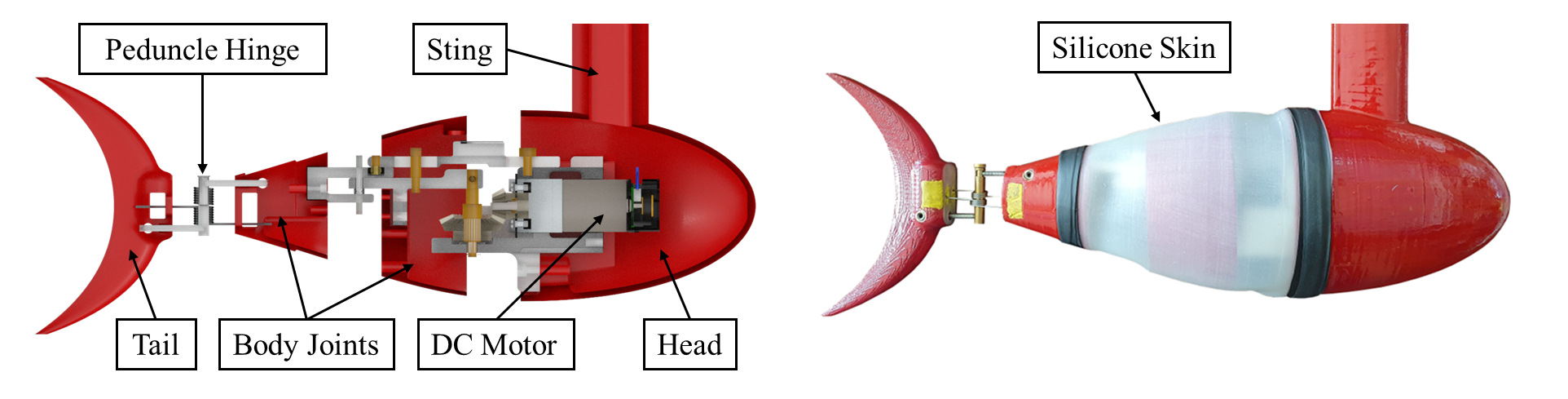}
\caption{CAD section of the fish robot in the left panel to present the internal components of the fish robot and the final prototype realization in the right panel.}
\label{fig:robot}
\end{figure}

\begin{figure}[h!]
\centering
\includegraphics[width=\linewidth,clip]
{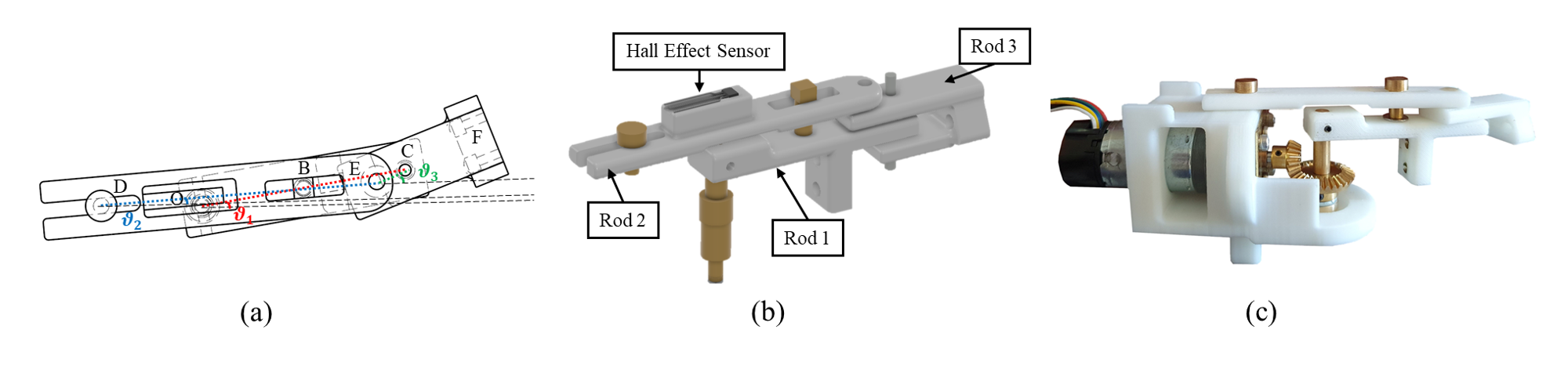}
\caption{Robot mechanism: Schematic representation (a), CAD rendering (b), final realization with assembled motor (c).}
\label{fig:mech}
\end{figure}

\begin{figure}[h!]
\centering
\includegraphics[width=\linewidth,clip]
{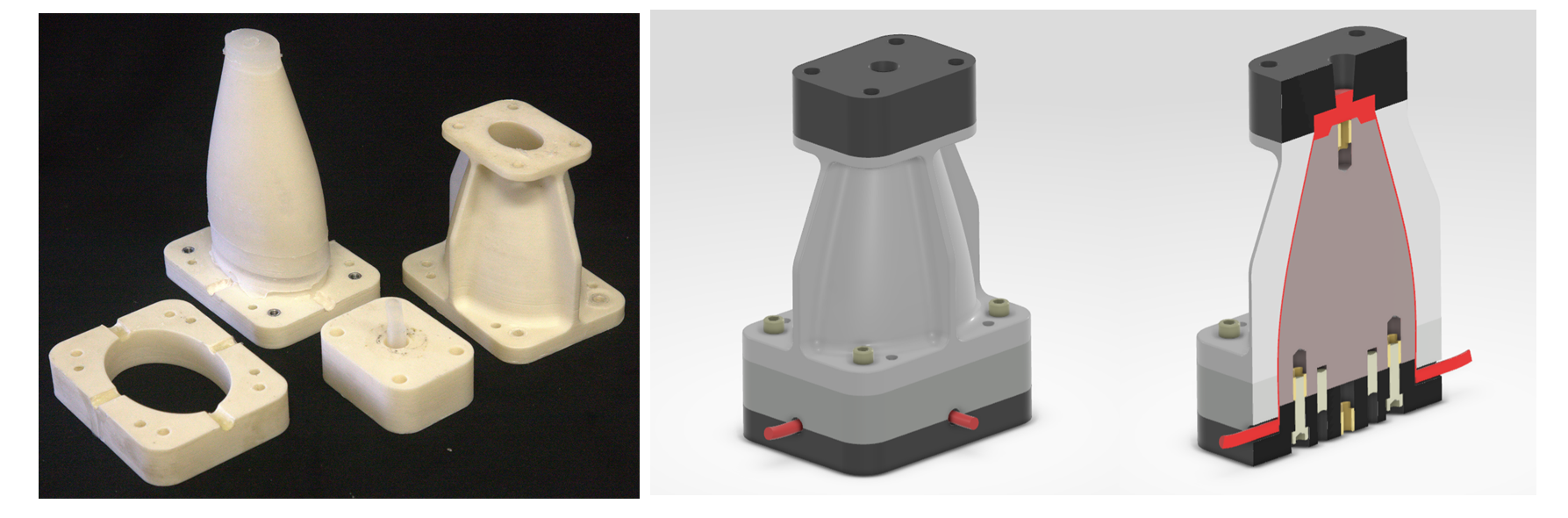}
\caption{Schematic and realization of the infusion mold used to create the silicone skin of the robotic fish. }
\label{fig:mold}
\end{figure}

\begin{figure}[h!]
\centering
\includegraphics[width=\linewidth,clip]
{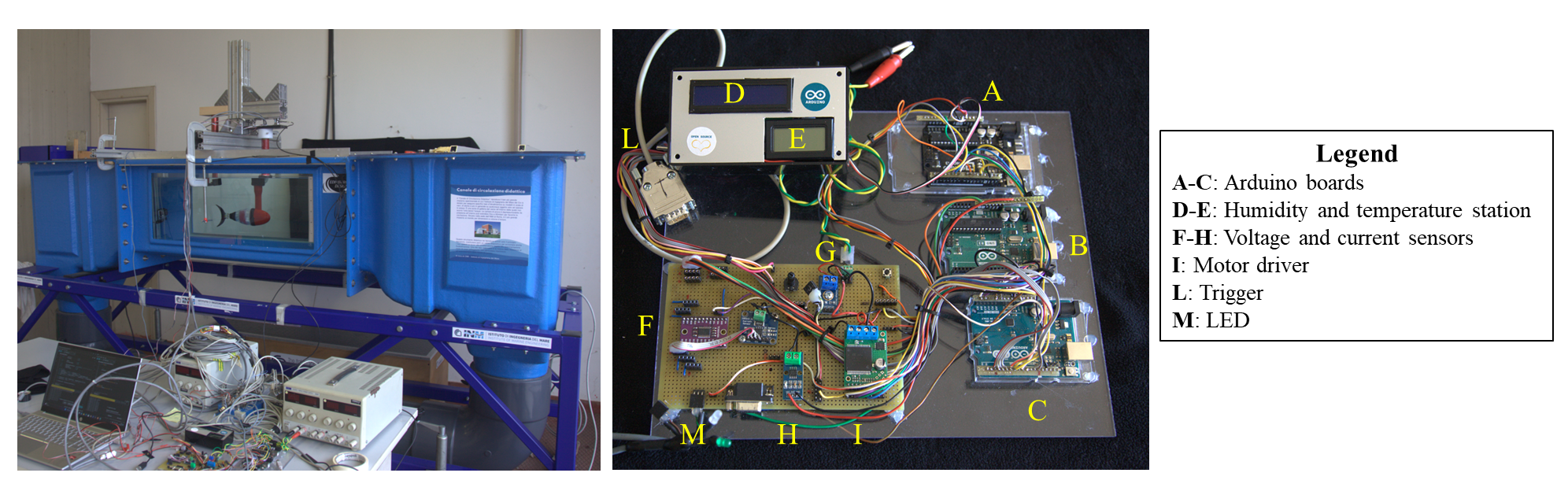}
\caption{Control electronics of the fish robot and control station (outside the recirculating channel).}
\label{fig:electronics}
\end{figure}

From a design point of view, our platform is inspired by a tuna of the Scombridae family whose configuration has been fully scanned by White et al. (\cite{article15}). A schematic representation is given by figure \ref{fig:robot}. 
\\The fish body, produced by a 3D printing machine, incorporates an internal DC motor capable of activating a mechanism with multiple joints connected to the tail through the peduncle hinge. Consequently, the tail performs a passive pitch motion, while the heave motion of the peduncle is controlled by the mechanism. \\
The tail is made passive using a pair of torsional springs. As successfully demonstrated that in aquatic propulsion, flexibility can be modelled using a concentrated spring at the peduncle (\cite{article4}). The spring has been chosen based on the estimated resonance frequency. To obtain the stiffness value $k$, a CAD estimate of the moment of inertia has been used, while the added mass value has been evaluated using estimates provided for ellipsoids (\cite{article17}):
\begin{equation}\label{eq:res_freq} 
f=\frac{1}{2 \pi} \sqrt{\frac{k}{ J_{yy}' +\lambda_{55}' } }
\end{equation}
where $f$ is the motion frequency and $(J_{yy}' +\lambda_{55}')$ is the sum of the moment of inertia and the added mass. Further details have been provided in the appendix. \\
Considering the heave motion of the peduncle, it depends exclusively on the mechanism. The mechanism consists of a series of rods connected to the DC motor shaft, where there is a pair of bevel gears to increase the torque output from the motor (see figure \ref{fig:mech}c).
\\The kinematics of the mechanism ensures that a sinusoidal motor input $\theta_1$ with amplitude $A$ and frequency $f$ results in an angular motion $\theta_3$ of the final joint proportional to the input angle according to a factor $\lambda$ dependent on the length of the rods:
\begin{equation} 
\begin{aligned} \theta_1 &= A \sin(2 \pi f t)  &\implies \theta_3 = \lambda (DE, OB, BC, EC, CF) \theta_1 
\end{aligned}
\end{equation}
This allows for a parabolic-like motion of the robot’s body segments, while the recovery of the traveling wave is managed by the tail. This mechanism has already been adopted for carangiform fish (\cite{article16}) and has proven to be particularly simple in its construction. For sake of conciseness, kinematic analysis has been reported in the appendix. \\
To guarantee waterproofing of the structure, coating treatments using epoxy resin have been applied. Additionally, between the various joints, a silicone rubber skin has been manufactured through a mold infusion process, represented in figure \ref{fig:mold}. To ensure a hydrodynamic shape and minimize friction, the silicone skin is 1 $mm$ thick and has been pretreated in a vacuum chamber to eliminate air bubbles. This way, the mixture is free of bubbles and ensures good impermeability even during tests, where the humidity inside the robot is monitored with a humidity sensor. The system has proven robust, maintaining a stable humidity value around 60 $\%$. The effect of humidity is however mitigated by the motor’s heating, which has been monitored using a thermocouple during testings.
\\The humidity and motor temperature values have been read in real-time through a station located outside the robot. As reported in Figure \ref{fig:electronics}, the control electronics is outside the robot. It consists of a circuit where Arduino boards and a driver control the motor via a position feedback system obtained through a Hall effect sensor mounted on one of the rods (see figure \ref{fig:mech}b). At the beginning of each experimental test, the position feedback allows for resetting with respect to the robot's symmetry plane, thus reducing the error in control at the start of each tail beat cycle. A LED is used to trigger all signals, including the cameras.  Additionally, current and voltage sensors have been used to measure the instantaneous consumption of the platform, acquiring data at a frequency of 500 $Hz$. Forces have been also acquired at a frequency of 500 $Hz$. These have been measured using piezoelectric force sensors placed outside the model and connected to the sting, to which the model has been attached. 
\\The sting is necessary to pass the motor and internal sensor cables and is cylindrical at the top while tapered at the bottom to reduce aerodynamic interference effects on the model. During measurements, the sting drag has been measured to obtain the resistance to be subtracted from the measurements taken on the model fish + sting configuration (see figure \ref{fig:setup}). Details on control and forces repeatability have been included in the appendix. \\
Finally, the entire system has been placed inside a recirculating channel whose speed can be modified and can reach 0.6 $m/s$ with a depth of 40 $cm$, as illustrated in figure \ref{fig:setup}. The platform is capable of oscillating in water within a frequency range of 1 to 3 $Hz$, with different amplitudes of the peduncle oscillation while the tail keeps to pitch passively. For this reason, video recordings have been necessary to obtain the robot's kinematics. The robot has been consequently painted black with a series of white markers, framed by a video system during the experiments. To make the kinematic measurements less intrusive, a mirror has been used under the model, which is captured from outside the channel by a camera positioned at a 45-degree angle, in order to reduce perspective distortion. Finally, the images have been processed in post-processing to reconstruct the motion of the passive fin.
\begin{figure}[h!]
\centering
\includegraphics[width=\linewidth,clip]
{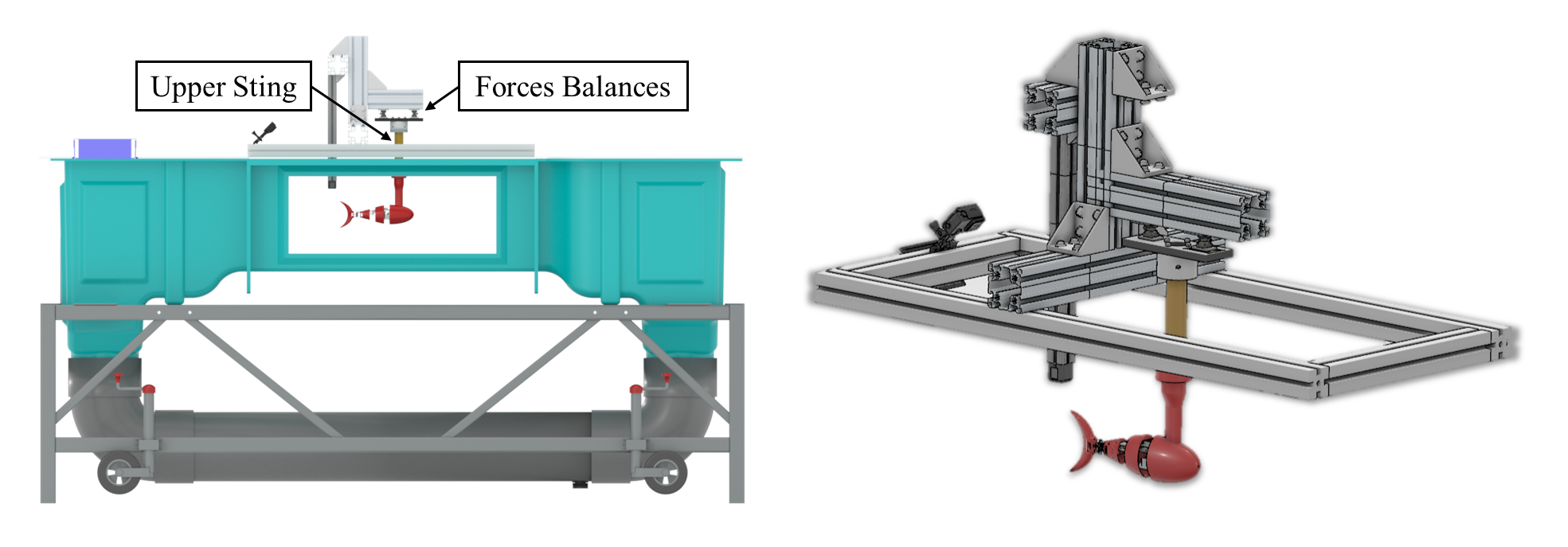}
\caption{Experimental setup for testing the robotic platform. Recirculating channel in the left panel and robot support in the right panel. }
\label{fig:setup}
\end{figure}

\section{Results and discussion}
\label{sec:results-discussion}
\begin{figure}[h!]
\centering
\subfigure[]{\includegraphics[viewport=100 300 510 500,width=0.32\linewidth,clip]{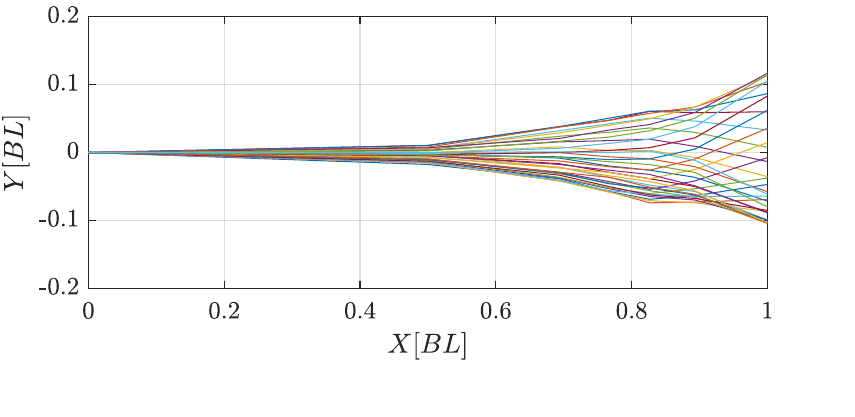}}
\subfigure[]{\includegraphics[viewport=100 300 510 500,width=0.32\linewidth,clip]{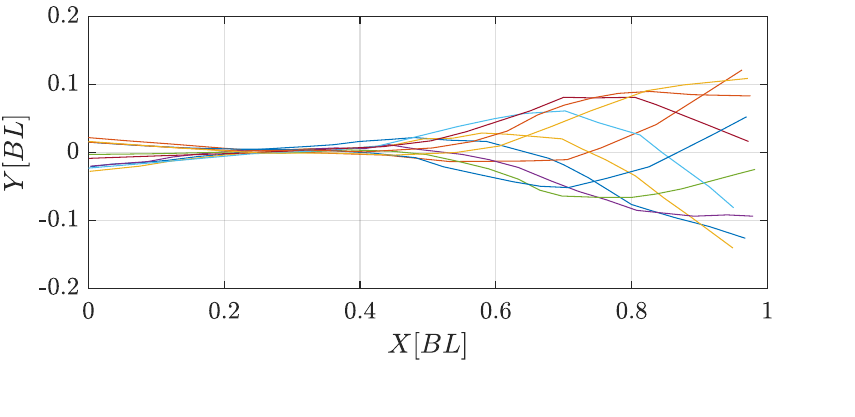}}
\subfigure[]{\includegraphics[viewport=100 300 510 500,width=0.32\linewidth,clip]{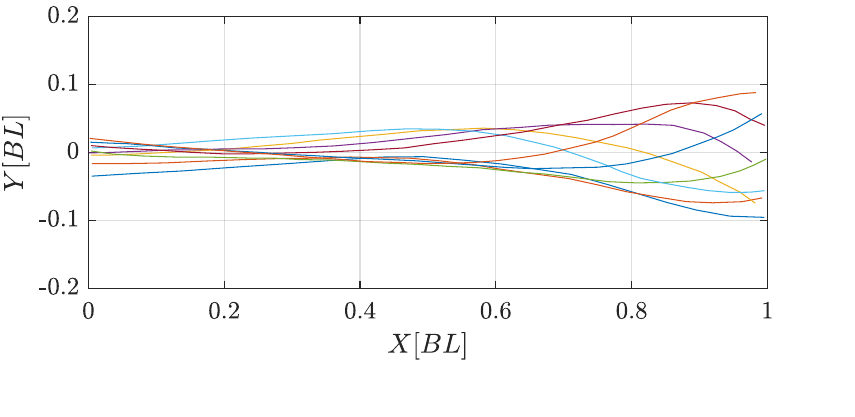}}
\caption{Midlines of the fish-like robot (a) measured during experiments compared with the midlines of the tuna-inspired robotic pltaform (b) and those of a tuna (c) (see \cite{article2},\cite{article15}).}
\label{fig:midlines_comparison}
\end{figure}

\begin{figure}[h!]
\centering
\subfigure[]{\includegraphics[viewport=100 260 510 580,width=0.45\linewidth,clip]{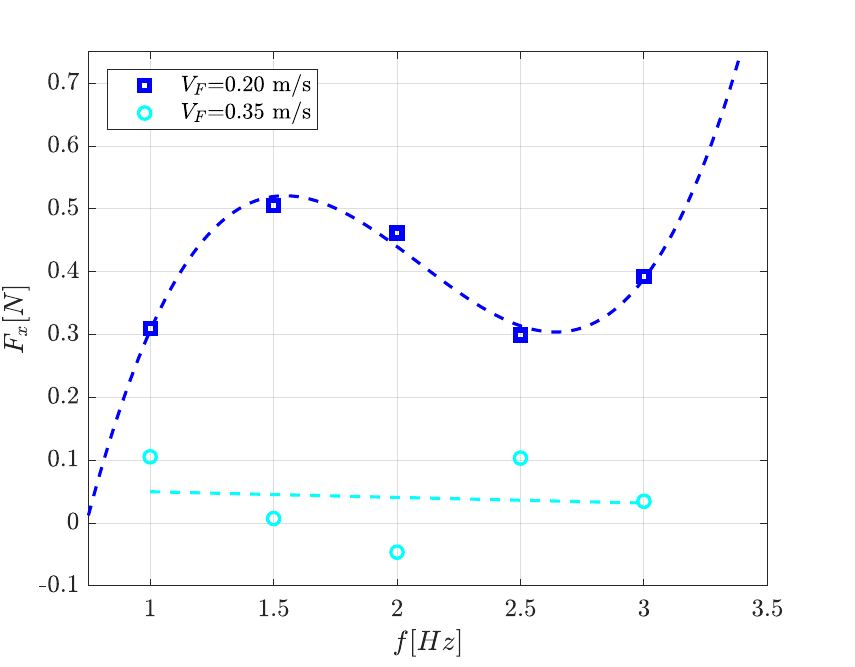}}
\subfigure[]{ \includegraphics[width=0.45\linewidth,clip]
{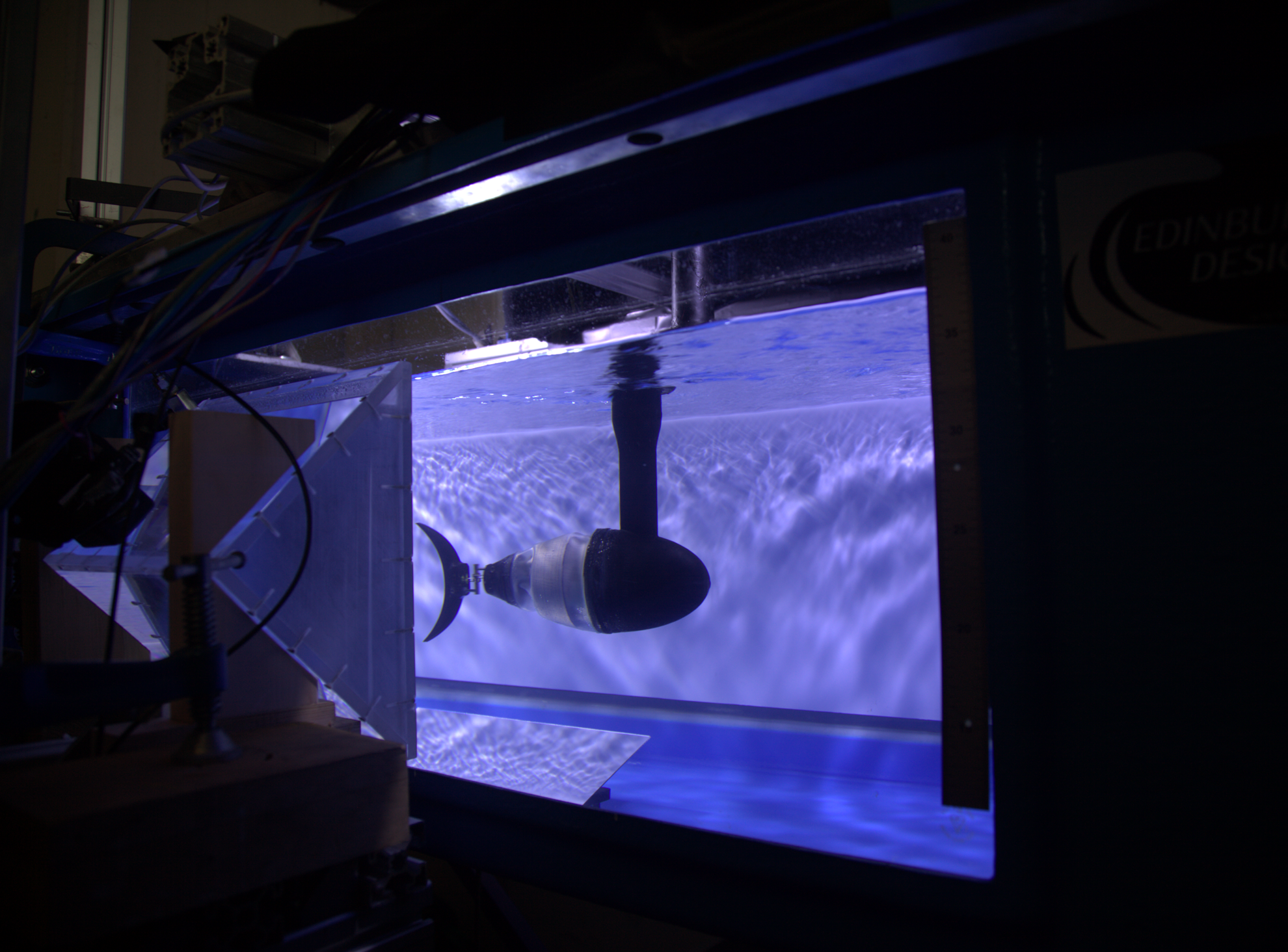}}
\caption{Axial force (a) as a function of frequency for different channel speeds and springs. Photo of the experiments (b).}
\label{fig:_Fx_springs}
\end{figure}

\begin{figure}[h!]
\centering
\subfigure[]{\includegraphics[viewport=100 260 510 580,width=0.32\linewidth,clip]{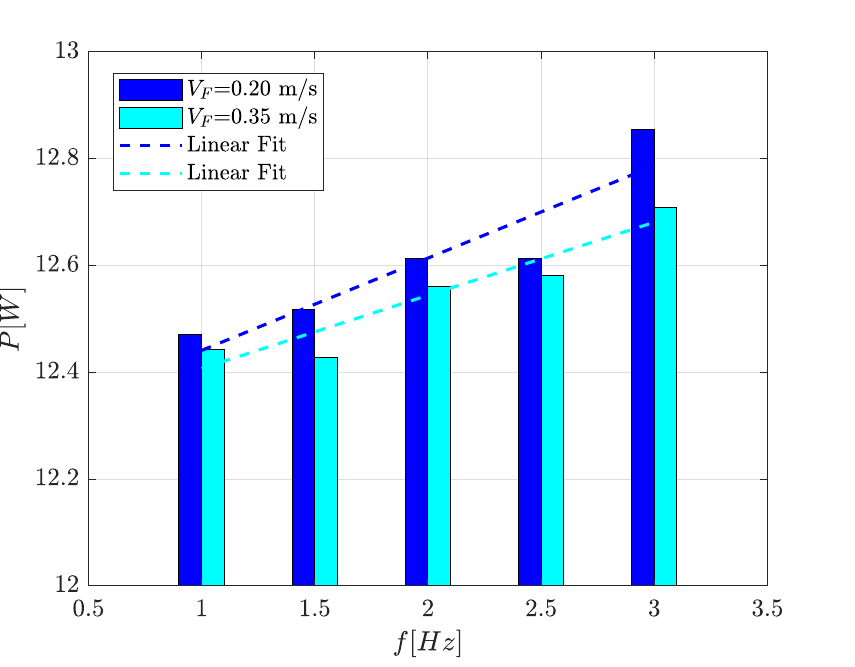}}
\subfigure[]{\includegraphics[viewport=100 260 510 580,width=0.32\linewidth,clip]{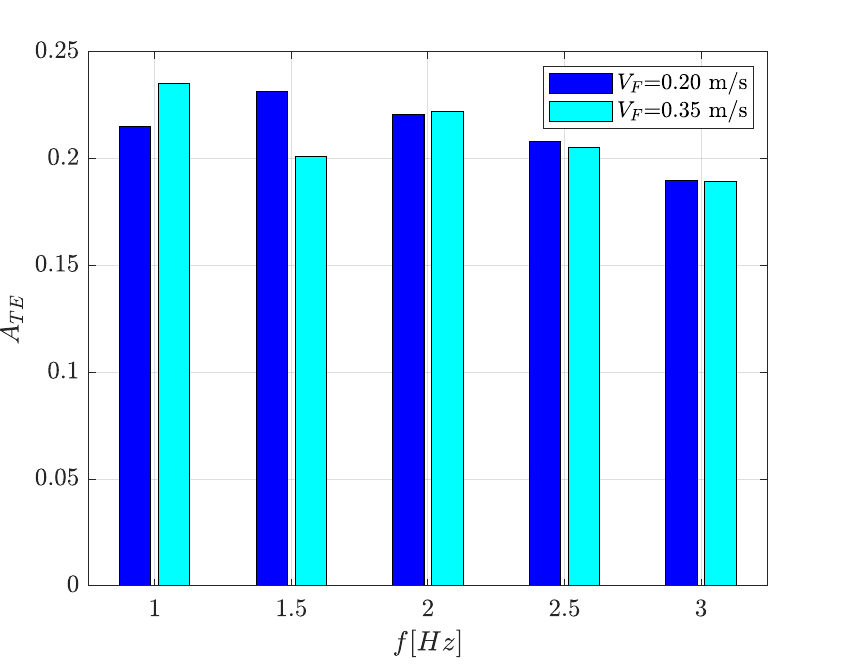}}
\subfigure[]{\includegraphics[viewport=100 260 510 580,width=0.32\linewidth,clip]{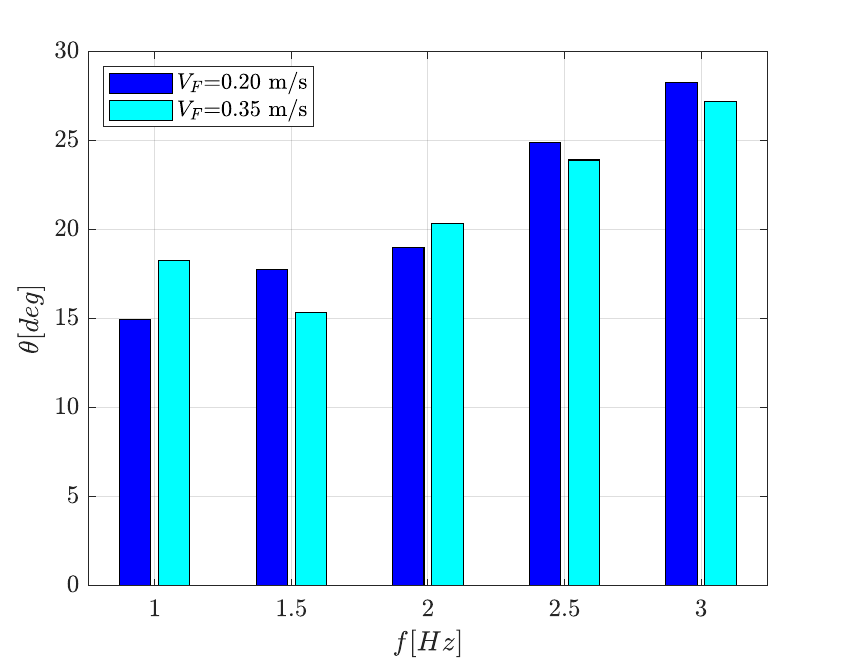}}
\caption{Power (a) as a function of frequency for different channel speeds and the same spring. Comparison of amplitude at the trailing edge (b) and pitch angle (c) as a function of frequency for two different channel speeds and the same spring.}
\label{fig:power_kinematics_same_spring}
\end{figure}

Figure \ref{fig:midlines_comparison}a depicts the midlines of the fish-robot, measured during a typical experiment through image reconstruction. Additionally, midlines are compared with another robotic tuna (fig. \ref{fig:midlines_comparison}b) and a real tuna (fig. \ref{fig:midlines_comparison}c). The experimental results exhibit a significant agreement in terms of maximum amplitude at the trailing edge among the different fish (both real and robotic). However, the real tuna displays a smoother deformation amplitude and a non-zero amplitude for $X$ approaching zero, indicating head recoil, which is absent in our case since the fish is attached to a sting. This figure anticipates how a passive fin can mimic the traveling wave deformation pattern typical of real fish. In other words, using a passive fin is enough to adjust the entire parabolic body motion, achieving the phase shift needed to create a traveling wave under high-efficiency conditions.
Figure \ref{fig:_Fx_springs} illustrates the results in terms of average force measured for the cases with different springs as function of robot frequencies $f$ for a single spring with stiffness 
$k$=58 $Nmm$ at different controlled channel speeds $V_F$ (see appendix for details on the relationship between the flow velocity and the controlled channel speed). Positive values are to be interpreted as thrust, while negative values indicate resistance. The dashed line represents the trend to better interpret the experimental data. \\
The curve seems to show a peak at 1.5 Hz. This peak could be related to resonance phenomena, although analytical predictions indicate a higher spring stiffness would be required to achieve resonance at 1.5 Hz. Some authors suggest that force is maximized under resonance conditions, before continuing to increase monotonically with frequency. To reinforce the earlier hypothesis we can analyze the kinematic quantities at speed $V_F$=0.20 $m/s$. As it can also be seen in \ref{fig:power_kinematics_same_spring}b, the trailing edge amplitude $A_{TE}$ is maximized under resonance conditions, as evidenced by several authors (\cite{article18}). 
\\We can then ask what happens when the channel speed is increased to $V_F$=0.35 $m/s$, using the same spring. The results in \ref{fig:_Fx_springs} prove that thrust indeed decreases, and the system condition evolves towards self-propulsion, where thrust equals drag. Clearly, the trend of the curve in cyan does not show exactly zero force $F_x$ but its average value is around 5 $g$, which can be considered zero given the accuracy of the force balance. Discovering the self-propulsion conditions finally allows the estimation of the robot's self-propulsion speed, which is 1.2 $BL$ with Strohual number equal to 0.35. Figure \ref{fig:power_kinematics_same_spring} offers insights for comparing a case with incoming flow and a self-propelled one in terms of power and kinematics. The trends in panel (a) highlight that power increases linearly with frequency, with lower energy consumption under self-propulsion conditions, as documented in the literature (\cite{article19}). 
\\When comparing the amplitude $A_{TE}$ in panel (b), it is possible to see that, on average, the value in the case of incoming flow (i.e. $V_F$=0.20 $m/s$) is higher than in the self-propulsion case, as the fish maximizes performance. Finally, in panel (c), the passive pitch angle $\theta$, derived from image reconstruction is reported . The value of $\theta$ on average increases with frequency, with higher values in the case of $V_F$=0.20 $m/s$. However, it can also be seen that the trends do not increase monotonically but tend to plateau, as observed also in analytical cases (\cite{article4}). 

\section*{Appendix}
\appendix
\subsection{Tail stiffness model}
The stiffness $k$ of the torsional spring for a motion at frequency $f$ is  derived from the equation:
\begin{equation}\label{eq:k}
k=(J_{yy}'+\lambda_{55}')(2\pi f)^2
\end{equation}
where $J_{yy}'$ and $\lambda_{55}'$ are the moment of inertia and the added mass of a slender ellipsoid with respect to the axis passing through its center and in the direction opposite to gravity. To find them, let us consider an ellipsoid with dimensions ($a$, $b$, $c$), made of PLA material with density $\rho_{PLA}$, mass $m$ and moment of inertia $J_{yy}$ with respect to the $y$-axis:
\begin{equation}\label{eq:mass_inertia}
\begin{aligned}
m &= \frac{4}{3}\pi\rho_{PLA} abc \\
J_{yy} &= \frac{4}{15}\pi\rho_{\small{PLA}} abc(a^2+c^2)
\end{aligned}
\end{equation}
Applying the Huygens-Steiner theorem, the moment of inertia $J_{yy}'$ with respect to an axis $y'$ located a distance $a$ from $y$ is given by:
\begin{equation}\label{eq:huyghens}
J_{yy}'=J_{yy}+ma^2
\end{equation}
Similarly, the added mass of the ellipsoid $\lambda_{55}'$ with respect to an axis $y'$ located a distance $a$ from $y$ (\cite{article17})  can be derived as:
\begin{equation}\label{eq:lambda55primo}
\lambda_{55}'=\lambda_{55}+\lambda_{33}a^2
\end{equation}
Where the added mass $\lambda_{ii}$ can be expressed as:
\begin{equation}
\begin{aligned}
\lambda_{33} &= k_{33} \bar{m} \\
\lambda_{55} &= k_{55} \bar{J_{yy}}
\end{aligned}
\end{equation}
With $\bar{m}$ and $\bar{J_{yy}}$ calculated as in eq. (\ref{eq:mass_inertia}), but with the density of the fluid, while the coefficients $k_{ii}$ are geometric functions of ($a$, $b$, $c$). The provided estimate suggests that to achieve a resonance frequency below 1.5 $Hz$, the stiffness value must be below 100 $Nmm$. This is a purely theoretical estimate with significant approximations, as the added mass calculation considered a body with uniform density equal to the printing material's density. Thus, a deviation in the added mass calculation could change the estimated frequency value.

\subsection{Mechanism kinematic analysis}

By performing kinematic analysis, we can determine the relationships between the links of the tail mechanism and the reference frame positioned at point O using the following expressions:
\begin{equation}
DO - DB\cos\theta_2 = -OB\cos\theta_1 \label{eq:1}
\end{equation}
\begin{equation}
DB\sin\theta_2 = OB\sin\theta_1 \label{eq:2}
\end{equation}
\begin{equation}
EC\cos\theta_3 = DO - DC\cos\theta_2 + OC\cos\theta_1 \label{eq:3}
\end{equation}
\begin{equation}
EC\sin\theta_3 = -DE\sin\theta_2 + OC\sin\theta_1 \label{eq:4}
\end{equation}
Solving the previous expressions, four unknown parameters, $DB$, $DO$, $\theta_2$ and $\theta_3$, can be obtained. Considering the actuation angle $\theta_1$ provided by means a DC motor as a function of the amplitude $A$, frequency $f$, and time $t$:
\begin{equation}
\theta_1 = A\sin(2\pi ft) \label{eq:5}
\end{equation}
the output of the kinematic analysis is;
\begin{equation}
\theta_3 = \lambda\theta_1 \label{eq:6}
\end{equation}
where $\lambda$ depends upon the sizes of the links of the tail mechanism, including $DE$, $OB$, $BC$, $EC$ and $CF$.

\subsection{Recirculating channel characterization}
\begin{figure}[h!]
\centering
\subfigure[]{\includegraphics[viewport=100 260 510 580,width=0.49\linewidth,clip]{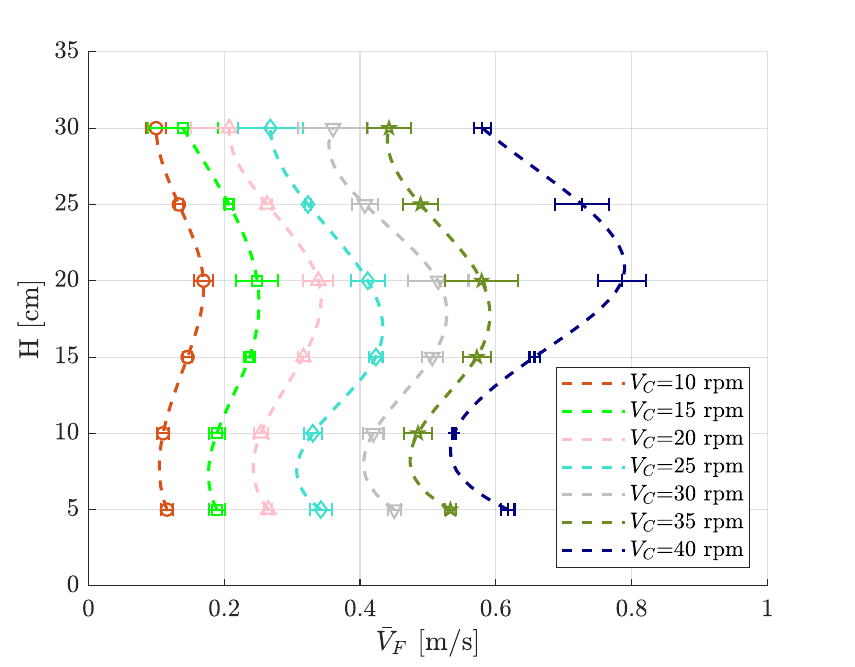}}
\subfigure[]{\includegraphics[viewport=100 260 510 580,width=0.49\linewidth,clip]{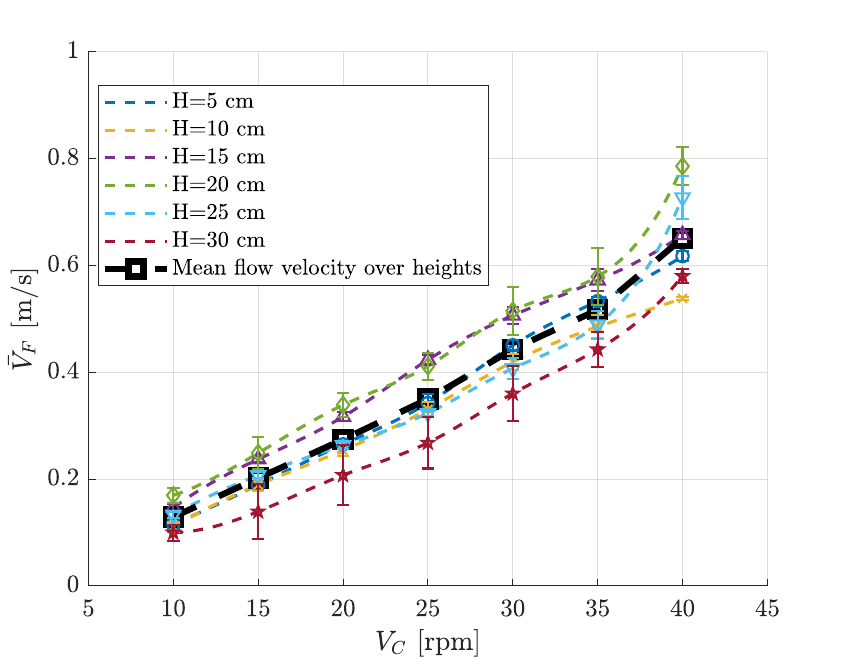}}
\caption{Flow speed profiles (a) as a function of height for different heights and tare curve (b) between channel and flow speeds.}
\label{fig:pitot}
\end{figure}

A Pitot tube, mounted at various longitudinal stations from the entering section (30-45-60 $cm$) and adjusted vertically to measure velocity profiles, has been used  to assess the influence of the free surface of the water, the tank floor, and flow uniformity. The channel is equipped with a control knob that regulates the angular velocity ($rpm$) of the propeller, which circulates water downstream of the exit section, achieving speeds of up to 1 $m/s$ in the test section. However, at high speeds, the flow becomes very non-uniform, and the honeycomb grid upstream of the test section is ineffective. For this reason, the analysis has been limited to a maximum speed of 0.6 $m/s$, which we consider acceptable for simulating the speed encountered by a 0.3 $m$ long robotic fish.  \\
This is illustrated in figure \ref{fig:pitot}a, where the velocity profiles for different speeds and different heights are presented, along with the standard deviation obtained by averaging the profiles along the three distances. \\
The calibration curve for the channel is finally reported in figure \ref{fig:pitot}b, averaging not only across different longitudinal positions of the Pitot tube but also across different heights, providing a reference for setting the $rpm$ for subsequent tests. Given the flow conditions, it has been decided to position the test object in that area, with the fish nose at a nominal height of 25 $cm$ from the bottom side of the channel.

\subsection{Experiments repeatability}
\begin{figure}[h!]
\centering
\subfigure[]{\includegraphics[viewport=100 260 510 580,width=0.49\linewidth,clip]{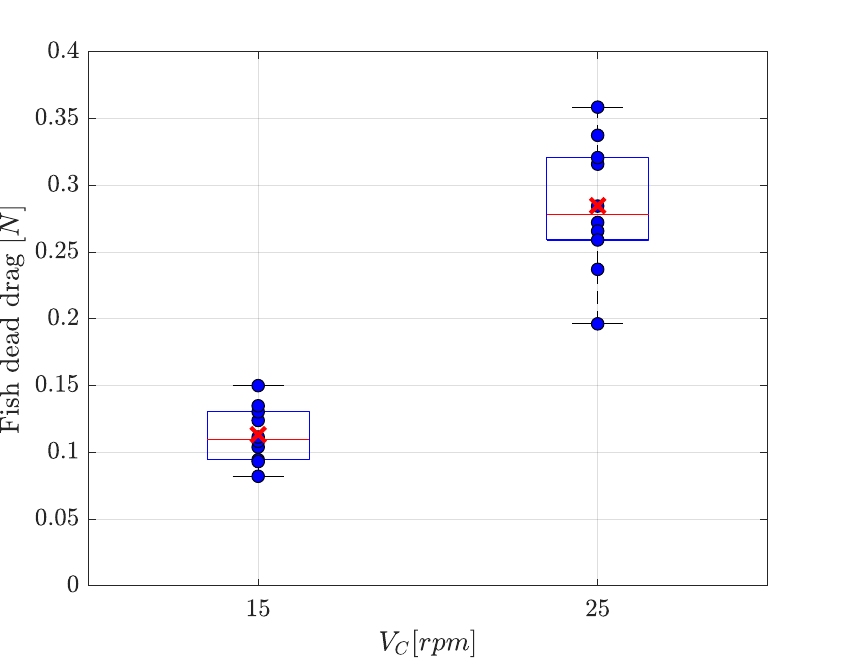}}
\subfigure[]{\includegraphics[viewport=100 260 510 580,width=0.49\linewidth,clip]{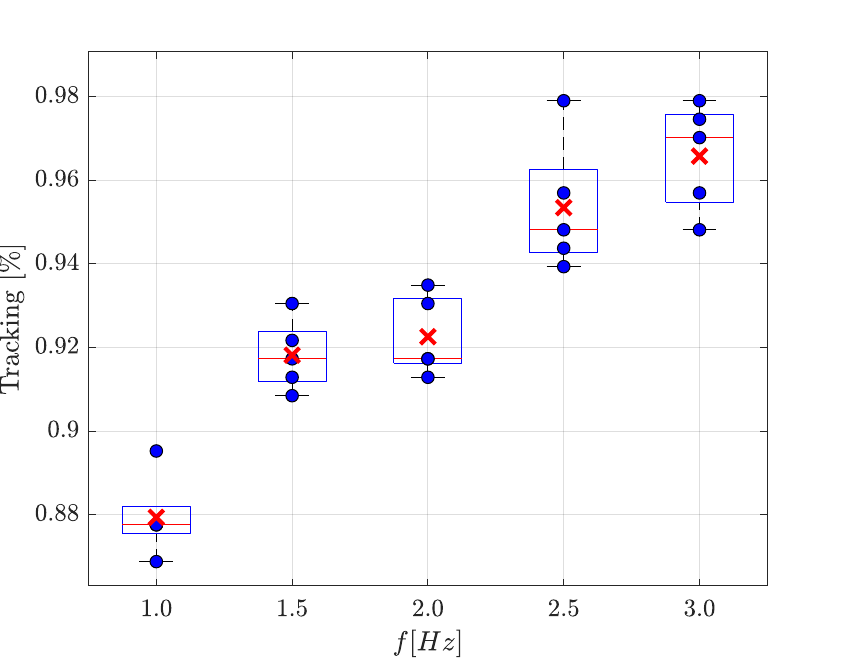}}
\caption{Fish dead drag (a) as function of channel speed for 10 experimental tests. Tracking percentage (b) as function of frequency for 5 trails. }
\label{fig:forces_tracking_repeat}
\end{figure}

\begin{figure}[h!]
\centering
\subfigure[]{\includegraphics[viewport=100 260 510 580,width=0.49\linewidth,clip]{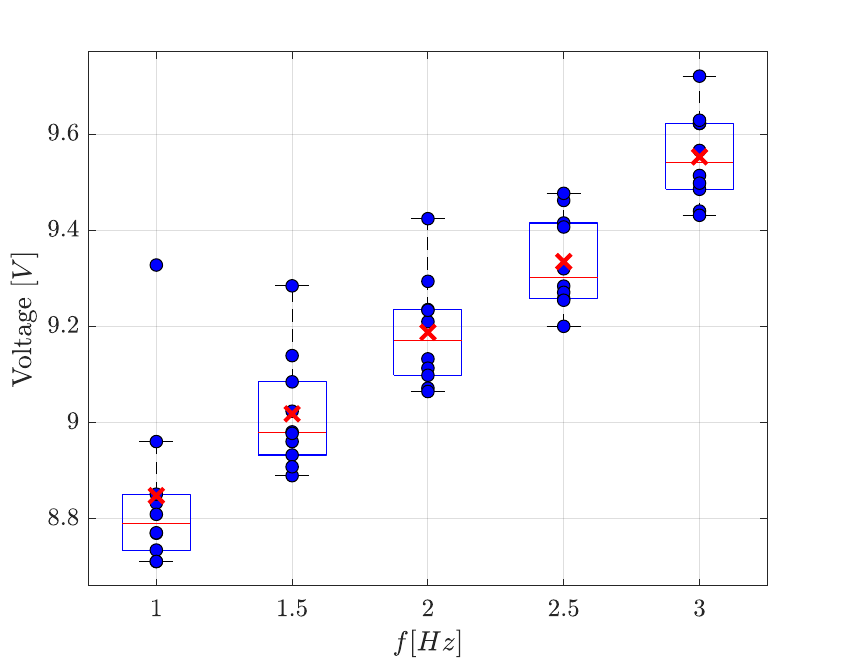}}
\subfigure[]{\includegraphics[viewport=100 260 510 580,width=0.49\linewidth,clip]{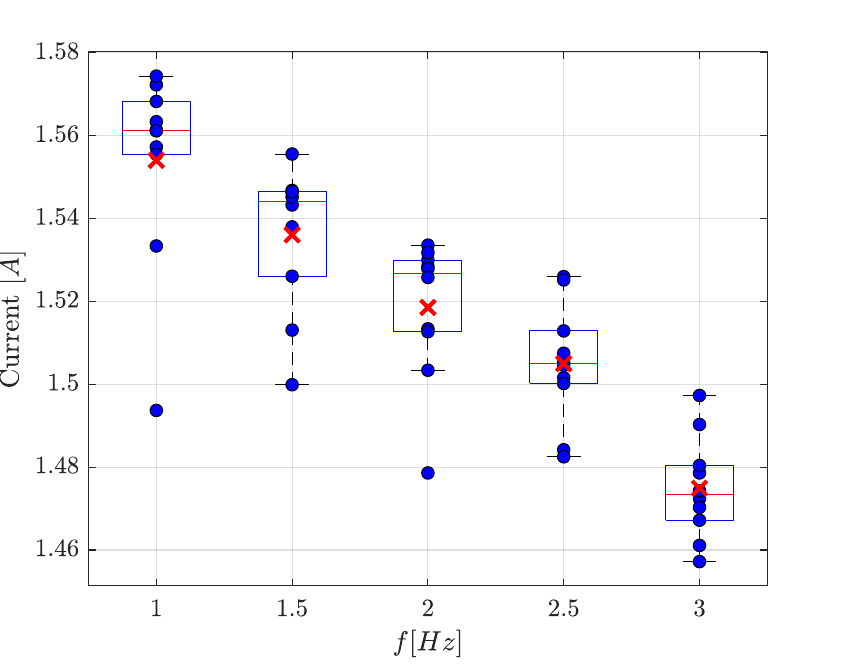}}
\caption{Voltage (a) and current (b) as function of frequency for 10 measurements.}
\label{fig:power_repeat}
\end{figure}

Figures \ref{fig:forces_tracking_repeat}-\ref{fig:power_repeat} present repeatability results for forces, electrical quantities, and control output. 
\\Figure \ref{fig:forces_tracking_repeat}a shows the repeatability analysis for fish dead drag (i.e. motionless fish drag) across 10 trials, demonstrating how dispersion increases at higher channel speeds. However, as observed from the experiments in section \ref{sec:results-discussion}, we have analysed a scenario at higher speeds only in one case, which is the case of self-propulsion (see figure \ref{fig:_Fx_springs}). \\
From control point of view, the repeatability has been quantified using the the tracking error, which describes how much different is the measured motion from the target. A value equal to 1 indicates perfect tracking. The error made in tracking a sinusoidal motion via a feedback system is presented for 5 trials in figure \ref{fig:forces_tracking_repeat}b, illustrating how the error is contained, slightly decreasing with frequency and is well repeatable. \\
Finally, regarding current and voltage, figure \ref{fig:power_repeat} indicates that voltage increases with the tail frequency, while current decreases. Across 10 tests, the analysis confirms good repeatability. 

\newpage
\section*{References}





\end{document}